\documentclass[unsortedaddress,prd,reprint,showpacs]{revtex4-1}

\usepackage{amsmath}
\usepackage{amsfonts}
\usepackage{amsthm}
\usepackage{amssymb}
\usepackage{graphicx}
\usepackage{hyperref}

	\newcommand{\beq}{\begin{equation}}
	\newcommand{\eeq}{\end{equation}}

	\renewcommand{\d}{\mathrm{d}}

\begin{document}

	\title{Thermodynamic curvature and ensemble nonequivalence}
	\date{\today}

	\author{Alessandro Bravetti}
        \email{bravetti@icranet.org}
	\affiliation{Instituto de Ciencias Nucleares, Universidad Nacional Aut\'onoma de M\'exico,\\ AP 70543, M\'exico, DF 04510, Mexico.}

	\author{Francisco Nettel}
	\email{Francisco.Nettel@roma1.infn.it}
	\affiliation{Dipartimento di Fisica, Universit\`a di Roma La Sapienza,\\ P.le Aldo Moro 5, I-00185 Rome, Italy}

	\begin{abstract}
In this work we consider thermodynamic geometries defined as Hessians of different potentials and derive some useful formulae that show their complementary role in the description of thermodynamic systems with two degrees of freedom that show ensemble nonequivalence. From the expressions derived for the metrics, we can obtain the curvature scalars in a very simple and compact form. We explain here the reason why each curvature scalar diverges over the line of divergence of one of the specific heats. This application is of  special interest in the study of changes of stability in black holes as defined by Davies. From these results we are able to prove on a general footing a conjecture first formulated by Liu, L\"u, Luo and Shao stating that different Hessian metrics can correspond to different behaviors in the various ensembles. We study the case of two thermodynamic dimensions. Moreover, comparing our result with the more standard turning point method developed by Poincar\'e, we obtain that the divergence of the scalar curvature of the Hessian metric of one potential exactly matches the change of stability in the corresponding ensemble. 	
	\end{abstract}


\maketitle

\section{Introduction}
In recent years there has been a lot of debate about the application of thermodynamic geometry to the investigation of the thermodynamic properties of black holes. 
In particular, many authors have been discussing the relationship between divergences of the thermodynamic curvature of 
different thermodynamic metrics and black holes phase transitions  (see, e.g., \cite{aman1,aman2, Rupp2007,Rupp2008, QuevGTDBHs,QuevSanchGTDAdSBHs, SSSindia,LiuLu,JankeJK,banarjee2011,sujoy2011,laszlo,HIGHERDIMBHS,GTDMPBH}).
Among these works, a major role is played by phase transitions as defined by Davies \cite{Davies}. Although the physical interpretation  of the lines individuated by Davies as real phase transitions
has been long discussed and there is no general consensus about the fact that a concrete change in the thermodynamic phase is present along such curves (see, e.g., \cite{Curir,Pavon,Lousto1,Lousto2,kaburaki1,kaburaki2,kaburaki3,Cai,ArcioniTelle}), 
however the divergence of second order response functions 
(most typically heat capacities) cannot be underestimated because it signals that some abrupt physical process occurs and instabilities develop. 

In this work, we will not enter into the discussion about the physical nature of Davies lines, but we will focus on a more geometrical aspect, that is, we will prove that the thermodynamic curvature defined
as the Hessian of one potential (say $F$) always diverges along lines of diverging heat capacity at some constant variable $X$ (i.e., $C_{X}$), while the thermodynamic curvature of the metric defined as the Hessian of its partial Legendre transform (say $M$)
always diverges at Davies lines of diverging $C_{Y}$, where $Y$ is the variable conjugated to $X$. 
This topic has so far represented a puzzle in the literature. An interesting insight about the nature of this correspondence has recently come from the work of Liu, L\"u, Luo and Shao \cite{LiuLu}. The main idea is that for systems like black holes, that show a different behavior depending on the ensemble that one is considering, it would be necessary to introduce more than one thermodynamic geometry, each one taking account of the characteristics of the ensemble corresponding to the potential used to define the metric.
Here we give both heuristic arguments and a general proof that the thermodynamic curvature of different Hessian metrics diverge along different Davies lines, at least for system with two degrees of freedom. This is the main result of the paper.

Moreover, comparing our results with the ones obtained by means of the turning point method developed by Poincar\'e \cite{kaburaki1,kaburaki2,kaburaki3,Cai,ArcioniTelle}, we can give strong further support to the idea that for systems showing ensemble nonequivalence (such as black holes) one needs to define different thermodynamic metrics, according to the different ensembles that can be individuated.

The structure of the paper is as follows: 
in Sec. \ref{sec:geometries} we introduce two different thermodynamic geometries based on the Hessians of different thermodynamic potentials and we derive useful mathematical formulae that highlight their complementarity. 
In Sec. \ref{sec:BHs} we present the general behavior of the thermodynamic curvatures along Davies lines for black holes. This is the main result of this work.  
Afterwards, we compare our results with the ones obtained using the turning point method and infer an interpretation of the role of thermodynamic curvature of Hessian metrics for systems with ensemble nonequivalence.
In Sec. \ref{sec:Conclusions} we review the  results and discuss some possible generalizations and future directions.
Finally in the Appendix \ref{appA} we present some examples derived from black holes thermodynamics.

\section{Thermodynamic geometries from different potentials}\label{sec:geometries}
\subsection{Thermodynamic metrics}
Consider a general thermodynamic system with two degrees of freedom described by a thermodynamic potential $M=M(S,X)$ where $S$ is the entropy and $X$ is the other control parameter for the system. In general, in the case of black hole thermodynamics the potential $M$ is the ADM mass and represents the internal energy of the system, but here we leave its name unspecified given that for black holes with a cosmological constant $\Lambda$ this thermodynamic potential is identified with the enthalpy in the extended phase space where $\Lambda$ is considered as a conjugated pressure (see, e.g. \cite{MannKubiznak}). We will refer to this representation as the $M$-potential. Now consider the following Hessian metric for the space of equilibrium states with parameters $X^i=\{S,X\}$
\beq  \label{mhessian}
g^M = M_{SS}\ \d S \otimes \d S + 2 M_{SX}\ \d S \otimes \d X + M_{XX}\ \d X \otimes \d X,
\eeq
where we use the notation $M_{X^i X^j}=\partial^2_{X^i X^j} M$ . It can be shown that such a metric can be re-written in a more compact form as \cite{Salamon}
\beq  \label{mgeneral}
g^M = \d T \otimes \d S + \d Y \otimes \d X,
\eeq
where $T=\partial_S M$ is the temperature of the system and $Y=\partial_X M$ is the conjugated variable to $X$. We will refer in the following to Eq. (\ref{mhessian}) as the \emph{expression for $g^M$ in the parameters space} and to Eq. (\ref{mgeneral}) as the \emph{general or coordinate free expression for $g^M$} \cite{Salamon}.

A third form in which we can write $g^M$ is considering the free energy $F=M-TS$. Such function $F(T,X)$ has different control parameters as its natural variables, the temperature $T$ and the variable $X$. 
It can be shown that in terms of these variables and of the function $F$ the metric $g^M$ takes a diagonal form
\beq  \label{mhessian2}
g^M = -F_{TT}\ \d T \otimes \d T + F_{XX}\ \d X \otimes \d X.
\eeq

Now introduce  a different metric defined as the Hessian of the free energy $F(T,X)$, that is,
\beq   \label{fhessian}
g^F = F_{TT}\ \d T \otimes \d T + 2 F_{TX}\ \d T \otimes \d X + F_{XX}\ \d X \otimes \d X.
\eeq
The above definition gives of course the expression for this metric in terms of its natural coordinates $\{T,X\}$.
As we did above for $g^M$, we can now rewrite $g^F$ in the coordinate free expression
\beq   \label{fgeneral}
g^F = - \d S \otimes \d T + \d X \otimes \d Y,
\eeq
and we can further give $g^F$ in terms of the the $M$-potential and its natural coordinates $\{S,X\}$ to obtain
\beq  \label{fhessian2}
g^F = -M_{SS}\ \d S \otimes \d S + M_{XX}\ \d X \otimes \d X.
\eeq

The metrics $g^M$ and $g^F$ described by the sets of equations (\ref{mhessian}-\ref{mhessian2}) and (\ref{fhessian}-\ref{fhessian2}), respectively, show an interesting parallelism. Indeed, written in their natural coordinates, they look as full Hessian metrics of their respective potentials, while each of them becomes diagonal when rewritten in terms of the transformed potential and its own natural coordinates. We will use this complementary property in the following to find simple expressions to write the determinants and curvature scalars of the two metrics and to investigate their physical meaning.

We conclude this section by noticing that in principle for a system with two thermodynamic degrees of freedom we could also have defined the Hessian metrics of the potentials $H=M-YX$ and $G=M-TS-YX$. However, it can be shown that the resulting metrics $g^H$ and $g^G$ are related to the ones that we are using here by the relations $g^H=-g^F$ and $g^G=-g^M$ \cite{Salamon}. 
Alternatively, we could have also worked with metrics defined as Hessian of the entropy and of its Legendre transforms, but again we would have obtained similar results, due to the fact that conformal relations always exist between the Hessian metrics of the energy potentials and the Hessian metrics of the entropy potentials \cite{LiuLu}. Therefore, for the case of two degrees of freedom it is sufficient to use only the metrics $g^M$ and $g^F$.

\subsection{Thermodynamic curvatures}
The general expression for the curvature scalar of a two dimensional space can be calculated by means of the following expression 
\begin{multline}   \label{dosR}
R = - \frac{1}{\sqrt{\text{det}(g)}} \bigg\{ \bigg(\frac{g_{11,2} - g_{12,1}}{\sqrt{\text{det}(g)}} \bigg)_{,2} + \bigg(\frac{g_{22,1} - g_{12,2}}{\sqrt{\text{det}(g)}} \bigg)_{,1}  \bigg\} \\
 - \frac{1}{2 \text{det}(g)^2} \,\text{det}(h),
\end{multline}
with 
\beq
h = 
\begin{pmatrix}
g_{11} && g_{12} && g_{22} \\
g_{11,1} && g_{12,1} && g_{22,1}  \\
g_{11,2} && g_{12,2} && g_{22,2}  
\end{pmatrix},
\eeq
where as usual $g_{ij}$ are the components of the metric $g$ in some coordinate system and a comma stands for partial derivative with respect to the corresponding variable. Now, for the case of a Hessian metric of some function $\Phi$ in terms of its natural variables the expression \eqref{dosR} reduces to \cite{BrodyRivier}
\beq   \label{dosRhessian}
R_{\text{Hessian}} =   - \frac{1}{2 \text{det}(g)^2}\, \text{det}(h),
\eeq
with 
\beq
h = 
\begin{pmatrix}
\Phi_{,11} && \Phi_{,12} && \Phi_{,22} \\
\Phi_{,111} && \Phi_{,112} && \Phi_{,122}  \\
\Phi_{,112} && \Phi_{,122} && \Phi_{,222}  
\end{pmatrix}.
\eeq
On the other hand, for a diagonal metric (i.e., $g_{12}=0$), it is immediate to see that $\text{det}(h)=0$ and therefore Eq. (\ref{dosR}) can be also simplified to
\beq   \label{dosRdiagonal}
R_{\text{diagonal}} = - \frac{1}{\sqrt{\text{det}(g)}} \bigg\{ \bigg(\frac{g_{11,2}}{\sqrt{\text{det}(g)}} \bigg)_{,2} + \bigg(\frac{g_{22,1} }{\sqrt{\text{det}(g)}} \bigg)_{,1}  \bigg\}.
\eeq

Equations (\ref{dosRhessian}) and (\ref{dosRdiagonal}) provide us with two compact formulae to compute the curvature scalars of the thermodynamic metrics $g^M$ and $g^F$ defined above. Indeed, using Eqs. (\ref{mhessian}) and (\ref{mhessian2}), we can write $R^M$ explicitly in the following two ways 
\begin{widetext}
\begin{eqnarray}
R^M &=\frac{M_{SS}(M_{SXX}^2-M_{SSX}M_{XXX})+M_{XX}(M_{SSX}^2-M_{SXX}M_{SSS})
+M_{SX}(M_{SSS}M_{XXX}-M_{SSX}M_{SXX})}{2(M_{SS}M_{XX}-M_{SX}^2)^2}, \nonumber   \\
R^M &=\frac{-F_{TT}\,F_{TXX}^2+F_{XX}\,F_{TTX}^2+F_{TT}\,F_{TTX}\,F_{XXX}-F_{XX}\,F_{TXX}\,F_{TTT}}{2\,F_{TT}^2\,F_{XX}^2}, \label{curvatureM1}
\end{eqnarray}
\end{widetext}
while for $R^F$, using Eqs. (\ref{fhessian}) and (\ref{fhessian2}), we get 
\begin{widetext}
\begin{eqnarray}
R^F &=\frac{F_{TT}\,(F_{TXX}^2-F_{TTX}\,F_{XXX})+F_{XX}\,(F_{TTX}^2-F_{TXX}\,F_{TTT})
+F_{TX}\,(F_{TTT}\,F_{XXX}-F_{TTX}\,F_{TXX})}{2(F_{TT}\,F_{XX}-F_{TX}^2)^2}, \nonumber \\
R^F&=\frac{-M_{SS}\,M_{SXX}^2+M_{XX}\,M_{SSX}^2+M_{SS}\,M_{SSX}\,M_{XXX}-M_{XX}\,M_{SXX}\,M_{SSS}}{2\,M_{SS}^2\,M_{XX}^2}. \label{curvatureF1}
\end{eqnarray}
\end{widetext}
Again, these formulae for the curvature scalars reflect the complementarity between the two metrics $g^M$ and $g^F$. In the next section we will use these mathematical identities to study the relation between the divergence of some of the response functions and those of the thermodynamic curvatures and explain why this is interesting in the case of systems showing ensemble inequivalence, referring for simplicity to the case of black holes.

\section{Davies lines and thermodynamic geometries}\label{sec:BHs}
\subsection{Davies lines and second order response functions}
Consider a black hole of mass $M$. We can consider the black hole mass as our thermodynamic potential  $M(S,X)$. 
In this case the first law of black holes thermodynamics reads $\d M = T\d S + Y\d X$.%

We want to consider here Davies lines of divergence of some response functions \cite{Davies}. Therefore, we introduce the quantities
\begin{multline}  \label{responsefunctions}
C_X = T \,(\partial_T S)_X, \,\, C_Y = T\,(\partial_T S)_Y, \,\, \alpha = - X^{-1} (\partial_T X)_Y, \\
\kappa_T = - X^{-1} (\partial_Y X)_T, \,\, \kappa_S = - X^{-1} (\partial_Y X)_S, \quad  
\end{multline}
which are respectively the heat capacity at constant $X$, the heat capacity at constant $Y$, the rate at which the black hole acquires the quantity $X$ at constant $T$, the rate at which the black hole acquires $X$ at constant entropy and
the coefficient of thermal $X$. For instance, in the case of the Reissner-Nordstr\"om black hole the quantity $X$ and its conjugated variable $Y$ are the electric charge $X=Q$ and the electric potential $Y=\phi$, respectively, while for the Kerr black hole they are the angular momentum $J$ and the angular velocity $\Omega$ \cite{Davies}.

As it is known from standard thermodynamics \cite{Callen}, the second order response functions are related by the condition
\beq  \label{identity1}
C_Y - C_X = \frac{T X \alpha^2}{\kappa_T},
\eeq
which remains valid for non-additive and non-extensive systems such as black holes \cite{Davies}.
Moreover, along Davies lines, only one of the response functions (typically $C_X$) diverges, while $C_Y$, $\alpha$, $T$, and $X$ are continuous. This fact, together with the above-mentioned identity (\ref{identity1}),
lead to conclude that $\kappa_T$ vanishes at the Davies points. Therefore, we can see that the behavior at those points is
\beq  \label{kappatPT}
C_X|_{\text{Davies}} \sim  \left.\frac{1}{\kappa_T} \right|_{\text{Davies}},
\eeq
where the subscript ``Davies'' means that we are observing the limiting behavior along Davies curve.
Moreover, we can write another identity between the response functions, that is, 
\beq   \label{identity2}
\kappa_T - \kappa_S = \frac{T X \alpha^2}{C_Y}.
\eeq
Equations (\ref{identity1}) and (\ref{identity2}) imply a third identity which reads 
\beq  \label{identity3}
\frac{C_X}{C_Y} = \frac{\kappa_S}{\kappa_T}.
\eeq
Using the above identity (\ref{identity3}), it is immediate to see that along Davies lines, i.e., whenever $C_X$ ($C_Y$) diverges and all other quantities are finite, $\kappa_T$ ($\kappa_S$) vanishes.
This remark will be useful for the comments in the following subsection.

\subsection{Thermodynamic metrics and determinants in terms of the response functions}
Using the definitions of the response functions, Eq. (\ref{responsefunctions}), and Eq. (\ref{mhessian}), $g^M$ can be expressed equivalently in terms of the response functions as
\beq  \label{mresponsekt}
g^M = \frac{T}{C_X}\ \d S \otimes \d S - 2\frac{T\alpha}{C_X \kappa_T} \d S \otimes \d X + \frac{C_Y}{X \kappa_T C_X}\ \d X \otimes \d X,
\eeq
with determinant given by \cite{Santoro2004}
\beq  \label{detmresponsekt}
\text{det}(g^M) = \frac{T}{X \kappa_T C_X} = \frac{T}{X \kappa_S C_Y},
\eeq
where in the second equality we have used the identity (\ref{identity3}).

On the other hand, using Eq. (\ref{fhessian2}), we see that
\beq  \label{fresponsekt}
g^F = -\frac{T}{C_X}\ \d S \otimes \d S + \frac{C_Y}{X \kappa_T C_X}\ \d X \otimes \d X.
\eeq
with determinant 
\beq   \label{detfresponsekt}
\text{det}(g^F) = -\frac{T C_Y}{X \kappa_T C_X^2}= -\frac{T}{X \kappa_S C_X},
\eeq
where in the second equality we have used again the identity (\ref{identity3}).

We notice that the last equation can also be rewritten in a quite appealing form
\beq\label{detappealing}
\text{det}(g^F) = -\frac{C_{Y}}{C_{X}}\ \text{det}(g^{M})= -\frac{\kappa_{T}}{\kappa_{S}}\ \text{det}(g^{M})= -\gamma\ \text{det}(g^{M}),
\eeq
where in the first equality we have substituted Eq. (\ref{detmresponsekt}), while in the second one we have used the relation (\ref{identity3}), and in the last we have introduced the heat capacity ratio $\gamma\equiv C_{Y}/C_{X}$. 

From Eq. \eqref{kappatPT} and the first equality in (\ref{detmresponsekt}) we see that the determinant of $g^M$ at Davies curve remains finite and non-vanishing, 
\beq  \label{detmresponsektlimit}
\text{det}(g^M)|_{\text{Davies}} \sim \frac{T}{X}\bigg|_{\text{Davies}},
\eeq
while Eq. (\ref{detfresponsekt}) implies that $\text{det}(g^F)$ vanishes as $C_X$ diverges
\beq  \label{detfresponsektlimit}
\text{det}(g^F)|_{\text{Davies}} \sim \frac{1}{C_X}\bigg|_{\text{Davies}}.
\eeq
Accordingly, the situation is reversed along the line where $C_{Y}$ diverges, that is, $\text{det}(g^F)$ is finite and non-vanishing while $\text{det}(g^M)$ vanishes as $C_{Y}{}^{-1}$.

From the expressions for the curvature scalars (\ref{dosRhessian}) and (\ref{dosRdiagonal}), it is possible to convince oneself that as the determinant of the metric vanishes the curvature diverges (provided the other terms are sufficiently well behaved in the sense that they do not introduce any singularity nor a vanishing quantity which modifies the behavior of the determinant function at the Davies line dictated by the divergence of the heat capacity at constant $X$. Such is the case of black holes systems, as we will explicitly show in the next subsection). This heuristic observation, together with the limits derived in (\ref{detmresponsektlimit}) and (\ref{detfresponsektlimit}), lead us to suggest that there is a general mathematical reason according to which $R^{F}$ always diverges at Davies line where $C_{X}$ diverges, while $R^{M}$ is in general finite. On the contrary, along Davies line where $C_{Y}$ diverges the situation is reversed. In the next subsection we will give a proof of this statement for the case of black holes.

\subsection{Behavior of the curvature scalar for black holes at Davies lines}
Consider Eqs. (\ref{curvatureM1}) and (\ref{curvatureF1}), from where we see that the curvature scalars in fact depend only on the second and third derivatives of the potential. This implies that if we know a general form for the second derivatives, then we can compute the behavior of the curvature scalar for all the examples belonging to this general class.
Therefore, keeping in mind the case of black holes, we make the following remarks:
\begin{itemize}

\item[1)] We can single out the term $f(S,X)$ that provides the divergence in the denominator of $C_X$, i.e., we
can write
\beq
C_X=\frac{K(S,X)}{f(S,X)^n},
\eeq 
with $n\geq1$. Here we are assuming that the numerator $K(S,X)$ in the heat capacity at constant $X$ is a non-vanishing finite function in the region of thermodynamic interest. In this way Davies line is defined by $f(S,X)=0$ and its derivatives $\partial_{X^i}f(S,X)$ are non-vanishing finite functions at the Davies line. Using the definition of $C_X$, Eq. (\ref{responsefunctions}), we can always write 
\beq\label{MSSassumption}
M_{SS}=f(S,X)^n\,N(S,X),
\eeq 
where $N(S,X) = T/K(S,X)$ is also well-behaved and not vanishing along Davies line.

\item[2)] $M_{SS}$ goes to zero at Davies line and, depending on the value of $n$, its derivatives too. However, $M_{XX}$ and its first derivatives are finite on Davies line.
Therefore, in what follows we will write explicitly the behavior of the derivatives of $M_{SS}$ in terms of the function $f(S,X)$, while we will let the derivatives of $M_{XX}$ implicit, since they do not play a significant role. 
Using Eq. (\ref{MSSassumption}), we get
\begin{eqnarray}
M_{SSS}&=&n\,f^{n-1}\,f_S\,N+f^n\,N_S\label{MSSS},  \\
M_{SSX}&=&n\,f^{n-1}\,f_X\,N+f^n\,N_X\label{MSSX},
\end{eqnarray}
and the derivative of $N$ with respect to the entropy reads
\beq \label{Ns}
N_S = \frac{N^2}{T} \left( f^n - K_S \right).
\eeq
\end{itemize}
Therefore, plugging Eqs. (\ref{MSSassumption}-\ref{MSSX}) into the second equation in (\ref{curvatureF1}), we get 
\begin{widetext}
\begin{multline}\label{RFlimit}
R^F|_{\text{Davies}} \sim \frac{1}{2N^2 M_{XX}^2} \bigg[\left( n^2  f_X^2 N^2 f^{-2} + 2 n  f_X N N_X f^{-1}\right) M_{XX} \\ 
+ n f_X N^2 M_{XXX} f^{-1} - \left(n  f_S N f^{-n-1} + f^{-n} N_S\right) M_{SXX} M_{XX} -  N M_{SXX}^2 f^{-n} \bigg],
\end{multline}
\end{widetext}
whose general behavior can be described by the following relation
\beq\label{RFlimit2}
R^F|_{\text{Davies}} \sim a\ f^{-2} + b\ f^{-1} +c\ f^{-n} + d\ f^{-n-1},
\eeq
where in the above expressions we have omitted the terms that do not diverge in the $f=0$ limit and
$a$, $b$, $c$, and $d$ are functions of $S$ and $X$ which remain finite at the Davies line. Therefore, $R^F$ always diverges along Davies line.

Analogously, following the above remarks 1) and 2) and going through the same steps as for $R^F$, but using the first identity in Eq. (\ref{curvatureM1}), we get
\begin{widetext}
\beq\label{RMlimit}
R^M|_{\text{Davies}}\sim \frac{n\,f^{n-1}\,N}{2M_{SX}^{4}}\left\{M_{XX}\left[M_{SXX}\,f_{S}-n\,f^{n-1}\,N\,f_{X}\right]+M_{SX}\left[M_{SXX}\,f_{X}-M_{XXX}\,f_{S}\right]\right\},
\eeq
\end{widetext}
so that the general behavior in this case is
\beq\label{RMlimit2}
R^M|_{\text{Davies}} \sim {A\,f^{n-1}+B\,f^{2n-2}},
\eeq
where $A$ and $B$ are nonvanishing finite functions at the Davies line. This shows that $R^M$ is finite along Davies line $f=0$ for any $n\geq 1$ (for $n>1$ the finite terms that we have suppressed in the above equations will be the only ones surviving).
In the Appendix \ref{appA} we provide some explicit examples which comply with our assumption $M_{SS}=f(S,X)^n\,N(S,X)$ with $n\geq 1$ making the above statements more clear.
We remark that this mathematical treatment clarifies \emph{in general} what happens at Davies lines of diverging $C_{X}$, a proof that has been lacking in the literature for a long time.
Furthermore, repeating the same reasoning but at lines of diverging $C_{Y}$, one can show that the situation is reversed as expected. In fact $R^{M}$ always diverges along such lines while $R^{F}$ is finite in general.
In the next section we will also give a physical interpretation to this complementarity, based on the analogy with the results obtained for black holes by using the turning point method developed by Poincar\'e \cite{kaburaki1,kaburaki2,kaburaki3,Cai,ArcioniTelle}.

To conclude this subsection, we comment on the fact that in the Geometrothermodynamics (GTD) program there is a metric structure, dubbed $g^{II}$, that has been applied to describe black hole geometric thermodynamics and for which the curvature always diverge at the Davies line. We argue here that the reason for this behaviour resides in the fact that the metric $g^{II}$ (with the choice $\Phi=M$, see \cite{CRGTD}) is conformal to the metric $g^{F}$ used here, and the conformal factor does not play a significant role along Davies lines. 

Therefore, from this point of view the GTD metric $g^{II}$ is complementary to the Hessian metric $g^{M}$ in the description of Davies lines for black holes and in general for systems showing ensemble nonequivalence.
Moreover, in the GTD context it has also been found another thermodynamic metric which is conformal both to the metric $g^{M}$ and to the Hessian metric of the entropy. Such a metric was called the natural metric ($g^{\natural}$), because it has been argued that it is the metric which complies with the demands of GTD \cite{CRGTD} and it has been applied so far only to ordinary systems \cite{applicationsgnat}. 
We argue here that in the GTD context the metrics $g^{II}$ and $g^{\natural}$ can be regarded as two complementary metrics for describing different ensembles, as well as the two metrics $g^{F}$ and $g^{M}$ used here.

\subsection{Analogy with the Poincar\'e method}\label{sec:Poincare}

In standard thermodynamics the analysis of the local stability conditions is based on considerations on the Hessian of the relevant thermodynamic potential, and this implies conditions on the signs of the second order response functions \cite{Callen}. Such standard method is based on the hypothesis of the additivity of the entropy function. 
This is a correct assumption for ordinary systems, but in the case of black holes it is known that this is not the case \cite{Bardeen}. 
A direct consequence of the lack of the additivity of the entropy is therefore the fact that the changes in sign of the response functions do not necessarily represent a change in the stability, at least not in all the ensembles \cite{ArcioniTelle}.

Nevertheless,  Poincar\'e developed a simple method to identify changes of stability over a linear series of equilibrium configurations, which is based on the representation of the diagram of a thermodynamic variable, plotted against its conjugate, taken to be the control parameter of the system, the so-called conjugacy diagrams.
The strength of this method resides in the fact that it is not assumed the property of additivity for the total entropy of the system plus its surrounding, making this approach particularly suitable to study changes of stability in nonextensive systems.
The method, known as the ``turning point'' method, prescribes that there is a real change in the stability of the system in one ensemble only at those points where the corresponding conjugacy diagram shows a vertical tangent.
Due to its power and simplicity, the turning point method has been extensively used in the black holes literature, in particular to clarify the physical nature of the phenomenon occurring along Davies lines in different ensembles (see, e.g, \cite{kaburaki1,kaburaki2,kaburaki3,Cai,ArcioniTelle};
in particular, \cite{ArcioniTelle} contains a detailed description of the method, its advantages and drawbacks).

For our purposes here, it is remarkable that for the black holes studied in the literature, no change in stability happens at Davies line when $C_{X}$ diverges in the microcanonical ensemble according to the turning point method, while in the canonical ensemble there is a vertical tangent, i.e. a change in the stability in this ensemble. In fact, there is a striking correspondence with the results presented here, that is, the metric defined as the Hessian of $F$, representing the canonical ensemble in this analogy, has a curvature scalar that diverges exactly along Davies curve, while the metric defined as the Hessian of the $M$-potential, connected to the microcanonical ensemble, does not.

To conclude these lines about the analogy a remark is in order. Using the turning point method one can detect the lines of change of stability in each ensemble. Nevertheless, the method cannot assure that one of the two phases is completely stable. In fact, a change of stability can occur from an unstable phase to a ``more unstable'' one. Also, the presence of a bifurcation on the conjugacy diagram cannot be signalized by this method and the relationship with the development of dynamic instabilities is not fully understood \cite{ArcioniTelle}. 
Nevertheless, this method clarifies the discussion about the nature of Davies lines of black holes and it strikingly agrees with the information coming from the use of the thermodynamic geometries $g^{M}$ and $g^{F}$ presented here.

\section{Conclusions}\label{sec:Conclusions}

In this work we have analyzed the thermodynamic geometry of general thermodynamic systems with two degrees of freedom that show ensemble nonequivalence, referring for clarity to black holes. 
To do so, we have focused on the Hessian metrics of different thermodynamic potentials, the so-called $M$-potential ($M=M(S,X)$) and its associated free energy ($F=F(T,X)$). In fact, we have argued that the critical behavior of such metrics qualitatively reproduces the behavior of the vast range of thermodynamical metrics defined in the literature, that is, Weinhold and Ruppeiner metrics, the Liu-L\"u-Luo-Shao set of metrics, and the GTD metrics based on symmetry with respect to Legendre transformations. This is so because all such metrics are Hessians of a thermodynamic potential or they are conformal to some Hessian metric.
 
As a starting point, we have presented here some interesting mathematical relations between the metrics $g^{M}$ and $g^{F}$ that hint to the fact that there exists a complementarity in the role of these two metrics. Moreover, by writing the metrics and their determinants in terms of the response functions, we argued heuristically that such complementarity can be given a physical interpretation, that is, the two metrics represent fluctuations in different ensembles.
 
In order to give this argument a more rigorous proof, we have written also the curvature scalars of the two metrics with respect to different sets of coordinates, thus showing that the complementarity also persists at the level of the curvature, as expected. 
This fact, together with some general remarks about the mathematical properties of Davies lines of change of stability for black holes, enabled us to prove that the curvature scalar of the $F$-potential diverges along Davies line where $C_{X}$ diverges, whereas the curvature of the $M$-potential in general stays finite.
On the contrary, along Davies line of diverging $C_{Y}$ the situation is reversed: the curvature of the $M$-potential diverges, while the curvature of the $F$-potential stays finite.
This result coincides with a previous description based on the analysis of several examples performed in \cite{LiuLu}. However, to the best of our knowledge, a precise mathematical explanation underlying this behavior for the Hessian metrics was not present in the literature.

In addition, we have seen that our results are in strict agreement to the ones obtained using the Poincar\'e method of turning points, giving a strong indication that the Hessian metrics corresponding to different potentials can carry a detailed information about the nonequivalence of the ensembles. In this context, it would be interesting to further investigate in the future the role of the thermodynamic curvatures as compared to the turning point method. In particular, we would like to understand if the thermodynamic curvatures can add some information about stability or presence of bifurcations, which cannot be fully described by means of the Poincar\'e method only.

Moreover, although we have been working here only with diverging heat capacities, we expect that the results can be easily generalized to the lines of divergence of any of the response functions (in particular it is of interest to investigate the behavior at lines of divergence of the generalized compressibilities, or generalized moments of inertia, because they correspond to the spinodal curves for fluid systems). We also consider that it would be interesting to extend the analysis to the case of systems with three or more thermodynamic degrees of freedom (see \cite{LiuLu, GTDMPBH} for examples), to see if an analogous result can be stated. 
We thus believe that these results can contribute to clarify the debate about the nature of phase transitions and changes of stability
 in the context of thermodynamic geometry for systems showing nonequivalence of the different ensembles, such as black holes.

To conclude, we  mention that Eq. \eqref{detappealing} has not been fully exploited here, but we expect that this relation will be found to have a physical meaning, considering that for fluids the heat capacity ratio can be related to the internal degrees of freedom and to the speed of sound.
 
\section*{Acknowledgements}
A. B. wants to thank the A. Della Riccia Foundation (Florence, Italy) for support. F. N. acknowledges support from CONACYT grant No. 207934.
The authors are grateful to the GTD group and in particular to C. S. Lopez-Monsalvo for valuable suggestions.

\appendix
\section{Examples from black holes thermodynamics}\label{appA}
In this Appendix we provide two examples taken from black holes thermodynamics which are the model examples in the context of black holes with two thermodynamic degrees of freedom, namely, the Reissner-Nordstr\"om and Kerr black holes. We report them here just to exemplify the validity and generality of our assumption $M_{SS}=f(S,X)^n\,N(S,X)$ with $n\geq 1$ and to make the statements in Sec. \ref{sec:BHs} more clear.
For more examples we refer to \cite{LiuLu}, where the authors performed a detailed analysis of the relation between the diverging response functions and the divergence of the curvature scalars $R^{M}$ and $R^{F}$ using several examples.
 
\subsection{Example: the Reissner-Nordstr\"om black hole}
The $M$-potential for the Reissner-Nordstr\"om black hole in $4$ dimensions reads \cite{HIGHERDIMBHS}
\beq
M(S,Q)=\frac{\sqrt{S}}{2}\left(1+\frac{Q^2}{S}\right).
\eeq
Therefore,
\beq
M_{SS}=-\frac{S-3Q^2}{8S^{5/2}},
\eeq
and hence
\beq
f=S-3Q^2\qquad N=-\frac{1}{8S^{5/2}}\qquad n=1.
\eeq
In this case we expect from Eqs. (\ref{RFlimit}) and (\ref{RMlimit}) that $R^F$ diverges as $f^{-2}$ at Davies line, while $R^M$ is finite.
Indeed, a direct computation gives
\beq
R^F=\frac{4S^{3/2}}{f^2}\qquad R^M=\frac{2S^{3/2}}{(S-Q^2)^2}\,.
\eeq

\subsection{Example: the Kerr black hole}
The $M$-potential for the Kerr black hole in $4$ dimensions reads  \cite{HIGHERDIMBHS}
\beq
M(S,J)=\sqrt{\frac{S}{4}+\frac{J^{2}}{S}}\,.
\eeq
Therefore,
\beq
M_{SS}=-\frac{S^{4}-24S^{2}J^2-48J^{4}}{8S^{5/2}\,(S^{2}+4J^{2})^{3/2}}
\eeq
and hence
\begin{eqnarray}
f &= S^{4}-24S^{2}J^2-48J^{4}, \\ 
N &= -\frac{1}{8S^{5/2}\,(S^{2}+4J^{2})^{3/2}}, \,\, \text{and} \,\, n=1.
\end{eqnarray}
A direct calculation from Eqs. (\ref{RFlimit}) and (\ref{RMlimit}) shows that $R^F$ diverges as $f^{-2}$ at Davies line, while $R^M$ is finite
\beq
R^F={\frac {18 \left( {S}^{2}+4\,{J}^{2} \right) ^{7/2} \left( S^{2}-4\,J^{2}
 \right) }{{S}^{3/2} f^{2}}},
\qquad R^M=0\,.
\eeq


\begin{thebibliography}{99}

\bibitem{aman1} 
J. E. \AA man, I. Bengtsson and N. Pidokrajt, Gen. Relativ. Gravit. \textbf{38}, 1305 (2006).

\bibitem{aman2}
J. E. \AA man and N. Pidokrajt, Phys. Rev. D \textbf{73}, 024017 (2006).


\bibitem{Rupp2007}
G. Ruppeiner, Phys. Rev. D \textbf{75}, 024037 (2007). 

\bibitem{Rupp2008}
G. Ruppeiner, Phys. Rev. D \textbf{78}, 024016 (2008).


\bibitem{QuevGTDBHs}
H. Quevedo, Gen. Relativ. Gravit. \textbf{40}, 971 (2008).

\bibitem{QuevSanchGTDAdSBHs}
H. Quevedo and A. S‡nchez, J. High Energy Phys. \textbf{034}, 0809 (2008).

\bibitem{SSSindia}
A. Sahay, T. Sarkar and G. Sengupta, J.High Energy Phys. \textbf{082}, 1007 (2010).

\bibitem{LiuLu}
H. Liu, H. Lu, M. Luo, K. N. Shao, J.High Energy Phys. \textbf{054}, 1012 (2010). 

\bibitem{JankeJK}
W Janke, D A Johnston, and R Kenna, J. Phys. A, \textbf{43}(42), 425206 (2010).

\bibitem{banarjee2011}
R. Banerjee, S.  Ghosh and D. Roychowdhury, Phys. Lett. B \textbf{696}, 156 (2011). 

\bibitem{sujoy2011}
R. Banerjee, S. K. Modak, S. Samanta, Phys. Rev. D \textbf{84}, 064024 (2011).

\bibitem{laszlo}
L. A. Gergely, N. Pidokrajt and S. Winitzki, Eur. Phys. J. C \textbf{71}, 1569 (2011).

\bibitem{HIGHERDIMBHS}
A. Bravetti, D. Momeni, R. Myrzakulov and H. Quevedo, Gen. Relativ. Gravit. \textbf{45}(8), 1603 (2013).

\bibitem{GTDMPBH}
A. Bravetti, D. Momeni, R. Myrzakulov and A. Altaibayeva, Advances in High Energy Physics \textbf{2013}, 549808 (2013).

\bibitem{Davies}
P. C. W. Davies, Proc. R. Soc. London, Ser. A \textbf{353}(1675), 449 (1977).

\bibitem{Curir}
A. Curir, Gen. Relativ. Gravit. \textbf{13}, 417 (1981).

\bibitem{Pavon}
D. Pavon, Phys. Rev. D \textbf{43}, 2495 (1991).

\bibitem{Lousto1}
C. O. Lousto, Nucl. Phys. B \textbf{410}, 155 (1993); \textbf{449}, 433 (1995).[erratum]

\bibitem{Lousto2}
C. O. Lousto, Int. J. Mod. Phys. D \textbf{6}, 575 (1997).

\bibitem{kaburaki1}
J. Katz, I. Okamoto and O. Kaburaki, Class. Quant. Grav. \textbf{10}, 1323 (1993).

\bibitem{kaburaki2}
O. Kaburaki, I. Okamoto and J. Katz, Phys. Rev. D \textbf{47}, 2234 (1993). 

\bibitem{kaburaki3}
O. Kaburaki, Gen. Relativ. Gravit. \textbf{28}, 7 (1996).

\bibitem{Cai}
Rong-Gen Cai, Zhi-Jiang Lu and Yuan-Zhong Zhang, Phys. Rev. D \textbf{55}, 853 (1997).

\bibitem{ArcioniTelle} 
G. Arcioni and E. Lozano-Tellechea, Phys. Rev. D \textbf{72}, 104021 (2005). 

\bibitem{MannKubiznak}
N. Altamirano, D. Kubiznak, R. B. Mann, Z. Sherkatghanad, Galaxies \textbf{2}, 89 (2014).

\bibitem{Salamon}
L. Salamon, J. Nulton and E. Ihrig, J. Chem. Phys. \textbf{80}, 436 (1984).

\bibitem{BrodyRivier}
D. Brody and N. Rivier, Phys. Rev. E \textbf{51}, 1006 (1995). 

\bibitem{Callen}
H. B. Callen, 
\textit{Thermodynamics and an Introduction to Thermostatistics}, 
$2$nd edition, (John Wiley and Sons, 1985).

\bibitem{Santoro2004}
M. Santoro, J. Chem. Phys. \textbf{121}, 7, 2932 (2004).

\bibitem{CRGTD}
A. Bravetti, C. S. Lopez-Monsalvo, F. Nettel and H. Quevedo, J. Math. Phys. \textbf{54}, 033513 (2013).

\bibitem{applicationsgnat}
A. Bravetti, C. S. Lopez-Monsalvo, F. Nettel and H. Quevedo, J. Geom. Phys. \textbf{81}, 1 (2014).

\bibitem{Bardeen}
J. M. Bardeen, B. Carter and S. W. Hawking, Commun. Math. Phys. \textbf{31}, 161 (1973).










\end{thebibliography}
\end{document}